\documentstyle[12pt]{article}
\begin{document}

\newtheorem{theorem}{Proposition}
\begin{flushright}
RIBS-PH-5/97

solv-int/9703012
\end{flushright}

\begin{center}
{\large{\bf LAGRANGIAN DESCRIPTION, \\[2mm] SYMPLECTIC STRUCTURE, 
AND INVARIANTS \\[2mm] OF 3D FLUID FLOW }} \\ 
\vspace{1cm}
\large{Hasan G\"{u}mral}  \\  \vspace{1cm}

T\"{U}B\.ITAK \\
Research Institute for Basic Sciences  \\
P.O. Box 21, 41470 Gebze-Kocaeli, Turkey \\ 
e-mail:hasan@mam.gov.tr  \\

\end{center}

\vspace{25mm}

\section*{Abstract}

Three dimensional unsteady flow of fluids in the Lagrangian description
is considered as an autonomous dynamical system in four dimensions. 
The condition for the existence of a symplectic structure on 
the extended space is the frozen field equations 
of the Eulerian description of motion. 
Integral invariants of symplectic flow are related
to conservation laws of the dynamical equation.
A scheme generating infinite families of symmetries and 
invariants is presented.
For the Euler equations these invariants are shown to have a geometric 
origin in the description of flow as geodesic motion; they are also
interpreted in connection with the particle relabelling symmetry.

\section{Introduction}

In spite of the powerful analytical tools in the study of fluid dynamics,
it has been argued that the three-dimensional Euler equation
requires more than just better functional analysis. 
The need for some geometry, more specifically, the necessity of symplectic
geometry for tackling some open problems of Euler flow is 
stressed in \cite{fri}.

Symplectic techniques have been used, on one side, to cast
the Eulerian dynamical equations into Hamiltonian (Lie-Poisson) 
form$^{\cite{mr}}$ and, 
on the other side, the Lagrangian description of steady fluid motion 
in two and four dimensions has been shown to
have the structure of a completely (Liouville)
integrable Hamiltonian system$^{\cite{gk}}$. 

In this work we shall be concerned 
with the kinematical problem of unsteady fluid motion,
namely, the Hamiltonian structure of the velocity field 
in the Lagrangian description of motion. This will be an application
in covariant framework of the construction in \cite{hg96} of 
the Hamiltonian structure of time-dependent dynamical systems.
The characterization of unsteady three dimensional
flow of a fluid as a Hamiltonian system in the symplectic framework
will involve the dynamical equations for frozen-in fields, thereby
manifesting an interplay with the Eulerian description.

The covariant symplectic techniques will enable us to construct 
Lagrangian and Eulerian invariants as well as to elucidate the
connection between them. 
We shall show that an infinite family of invariants can be
constructed for any hydrodynamic system admitting a single
vectorial frozen-in field. For the Euler equation the geometric
origin of this family will be found in the description of flow
as a geodesic motion on an infinite dimensional group.
We shall argue that the associated infinitesimal generators
belong to the left Lie algebra of particle relabelling symmetry.

\section{Motion of Fluids in Three Dimensions}

\subsection{Kinematic Representation}

Let the open set $D \subset R^{3}$ be the domain occupied initially
by the fluid and 
$x(t=0)=x_{0} \in D$ be the initial position, i.e., a Lagrangian label.
For a fixed initial position $x_{0}$,
the Eulerian coordinates $x(t) = g_{t}(x_{0})$ define 
a smooth curve in $R^{3}$
describing the evolution of fluid particles. 
Thus, for each time $t \in R$, the orientation (or volume) 
preserving embedding 
$g_{t}:D \to R^{3}$ describes a
configuration of fluid. A flow is then a curve $t \mapsto g_{t}$ 
in the space of all such maps.
The time-dependent Eulerian (spatial) velocity field $u_{t}$
that generates $g_{t}$ is defined by
\begin{equation}
    {dx \over dt}={dg_{t}(x_{0}) \over dt}
            =(u_{t} \circ g_{t})(x_{0})=u(t,x)    \label{vel}
\end{equation}
where $u_{t} \circ g_{t}$ is the corresponding Lagrangian (material)
velocity field$^{\cite{mr}}$.

Since $g_{t}$ is volume preserving, $u_{t}(x)$ is a divergence-free vector
field over $R^{3}$ and eq(\ref{vel}) is a non-autonomous dynamical system
associated with it.
Equivalently, (\ref{vel}) can be represented as an autonomous system by
the vector field
\begin{equation}
  \partial_{t} + u(t,x)   \;,\;\;\;\;\; u={\bf u} \cdot \nabla \label{tvel}
\end{equation}
on the extended space $R \times R^{3}$.

The Lagrangian description of fluid motion is the study of the system
of non-autonomous ordinary differential equations (\ref{vel}), or, 
equivalently, its autonomous counterpart associated with (\ref{tvel}).
We shall be interested in its Hamiltonian formalism which 
will turn out to imply and be implied by dynamical equations
for frozen-in Eulerian fields.

\subsection{Dynamical Equations}

The dynamics of the the Eulerian velocity field ${\bf u}(t,{\bf x})$ 
is determined by Newton's law. 
For an ideal incompressible isentropic fluid in a 
bounded domain $D$ with normal ${\bf n}$ this is given by the
Euler equations
\begin{equation}
  {\partial {\bf u} \over \partial t} +
  {\bf u} \cdot \nabla {\bf u}= - \nabla ({1 \over \rho} P)   \label{euler}
\end{equation}
where $\rho$ is the density and $P$ is the pressure.
The additional conditions $\nabla \cdot {\bf u} =0$ and
${\bf u} \cdot {\bf n} |_{\partial D} =0$ on the velocity field         
characterize it as the generator of volume preserving diffeomorphisms.
The identity ${\bf u} \cdot \nabla {\bf u}  = 
{1 \over 2} \nabla | {\bf u}|^{2} - {\bf u} \times (\nabla \times {\bf u} )$
can be used to bring the equation (\ref{euler}) into the form
\begin{equation}
  {\partial {\bf u} \over \partial t} -
  {\bf u} \times (\nabla \times {\bf u}) = 
  - \nabla ({1 \over \rho} P + {1 \over 2} u^{2} )   \label{reuler}
\end{equation}
which we shall refer to as the rotational Euler equation.
In terms of the vorticity vector ${\bf w} \equiv \nabla \times {\bf u}$
the Euler equation becomes
\begin{equation}
    {\partial {\bf w} \over \partial t} - 
    \nabla \times ( {\bf u} \times {\bf w} ) =0    \label{weq}
\end{equation}
and for our purposes we shall assume the existence of a scalar 
field $h$ advected by the fluid
\begin{equation}
    {\partial h \over \partial t} +
     {\bf u} \cdot \nabla h =0     \label{he}
\end{equation}
which further satisfies ${\bf w} \cdot \nabla h \neq 0$. This 
latter condition does not affect the generic structure 
but is only necessary to present the results in covariant symplectic form.

\section{Symplectic Structure of Kinematic Description}

A symplectic structure$^{\cite{mr},\cite{olver}}$ 
on a manifold $N$ of even dimension $2n$ is defined by a closed,
non-degenerate two-form $\Omega$. It is exact if there exist a 
one-form $\theta$ such that $\Omega =- d \theta$.
Darboux's theorem quarantees the existence of 
local coordinates $(q^{i},p_{i})\; i=1,...,n$ 
in which $\Omega$ has the canonical form $dq^{i} \wedge dp_{i}$.
The $2n-$form $(-1)^{n} \Omega^{n}/n!$ is called the Liouville volume.
A vector field $X$ on $N$ is called Hamiltonian if there exists
a function $h$ on $N$ such that 
\begin{equation}
      i(X)(\Omega)=dh                 \label{seq}
\end{equation}
where $i(X)(\cdot)$ denotes the inner product with $X$.
The identity $i(X)(dh)=0$ which follows from (\ref{seq}) is 
the expression for conservation of $h$ under the flow of $X$.
With the correspondence (\ref{seq}) between functions and vector fields,
the Poisson bracket of functions on $N$ defined by
\begin{equation}
   \{ f,g \} = \Omega (X_{f},X_{g}) = \Omega^{-1}(df,dg)   \label{pobi}
\end{equation}
satisfies the conditions of bilinearity, skew-symmetry, the Jacobi identity
and the Leibniz rule.
This enables us to write the dynamical system associated with the
vector field $X_{h}$ in the form of Hamilton's equations
\begin{equation}
   {dx \over dt} = \{ x,h  \}        \;.      \label{heq}
\end{equation}

\begin{theorem}

(1) The suspended vector field $\partial_{t}+u$ on $R \times R^{3}$
is a Hamiltonian vector field with the symplectic two-form
\begin{equation}
  \Omega_{e} = - (\nabla h + {\bf u} \times {\bf w}) 
          \cdot d{\bf x} \wedge dt + {\bf w} \cdot 
             (d{\bf x} \wedge d{\bf x})    \label{symp}
\end{equation}
and the Hamiltonian function $h$. 

(2) The potential vorticity $- {\bf w} \cdot \nabla h \equiv \rho_{h}$ 
is the invariant Liouville volume density of symplectic structure. 

(3) 
$\Omega_{e}$ is exact with the one-form
\begin{equation}
  - \theta_{e} =  \psi_{e} \, dt  + {\bf u} \cdot d{\bf x} \;,\;\;\;\;              
      -\psi_{e} = h + {1 \over \rho} P + {1 \over 2} u^{2} \;.  \label{cone}
\end{equation}

(4) In Darboux coordinates,
$\Omega_{e} = dq \wedge dp + dt \wedge dh_{can}$ and the velocity field
has the Clebsch representation 
\begin{equation}
  {\bf u}= \nabla s  + p \nabla q
\end{equation}
where $s$ is the generating function of the canonical transformation.
\end{theorem}

{\bf Proof:}
$\Omega_{e}$ is closed via eq(\ref{weq}) and the divergence
free property of vorticity. 
The non-degeneracy is a consequence of the assumption that
$-{\bf w} \cdot \nabla h = \rho_{h} \neq 0$ where
\begin{equation}
 {1 \over 2} \Omega_{e} \wedge \Omega_{e} = 
 \rho_{h} \, dx \wedge dy \wedge dz \wedge dt           \label{lvol}
\end{equation}
is the definition of the Liouville volume.
Its invariance can either be checked directly by showing that
\begin{equation}
   div_{\rho_{h}}( \partial_{t}+u)=
  {\partial \rho_{h} \over \partial t} +
  \nabla \cdot ( \rho_{h} {\bf u} ) = 0  \;.               \label{divo}
\end{equation}
or more geometrically, follows from the recognition that
eq(\ref{weq}), when written
in terms of differential operators $u={\bf u} \cdot \nabla$ and
$\omega ={\bf w} \cdot \nabla$ as
\begin{equation}
    {\partial \omega \over \partial t} + [ u , \omega ] 
     = - (\nabla \cdot {\bf u}) \omega  \;,   \label{obeq}
\end{equation}
is the criterion for $\omega$ to be a time-dependent symmetry of $u$. 
Then, the fact that, for dynamical systems,
action of symmetries on conserved quantities gives conserved 
quantities$^{\cite{olver}}$ implies the invariance of $\rho_{h}$.
See also \cite{sagdeev} for a physical
description of Lagrangian invariants produced in this way.

Eq(\ref{he}) is used to write $\partial_{t}+u$ in Hamiltonian form.
The definition of vorticity field as a curl and the rotational
Euler equation (\ref{reuler}) turns $\Omega_{e}$ into an exact symplectic 
form with (\ref{cone}). 
Clebsch variables for velocity field as well as the form of 
Hamiltonian function 
\begin{equation}
   h_{can} = \psi_{e}  - {\partial s \over \partial t}
          - p {\partial q \over \partial t}
\end{equation}
in Darboux coordinates follows easily from the defining relations 
\begin{equation}
 ds=  \theta_{can} - \theta   \;,\;\;\;
  - \theta_{can}=pdq+h_{can}dt 
\end{equation}  
for these coordinates. $\bullet$

An important conclusion of this result is that the Eulerian dynamical
equations determining the form of velocity field enforce its
suspension to be a Hamiltonian vector field.
In the next section we shall show
that with a slight modification the physical framework of
proposition (1) can be extended to other hydrodynamic systems.

\section{Generalizations on Dynamical Equations}

The conditions for the existence
of Hamiltonian structure of the velocity field can be made
independent of the dynamics of the fluid itself by replacing
vorticity in $\Omega_{e} $ with a divergence free field
which is advected by fluid motion. This allows one to consider
fluid motion under more general forces.
Velocity fields governed by the Euler equation (\ref{euler}) has already
such a generalization which involves little alteration of their Hamiltonian
structure. Namely, any lift force can be incorparated
to the right hand side of eq(\ref{euler}) by the replacement
$\nabla h + {\bf u} \times {\bf w}  \mapsto 
\nabla h + {\bf u} \times {\bf w} + {\bf F}$ in $\Omega_{e}$.
The foremost example of this sort of force fields with vanishing 
component along ${\bf u}$ are the magnetic forces acting on 
electrically conducting fluids. 

\begin{theorem}
Let the dynamics of ${\bf u}$ be governed by 
\begin{equation}
  {\partial {\bf u} \over \partial t} +
  {\bf u} \cdot \nabla {\bf u}= {\bf F}   \label{geuler}
\end{equation}
and assume that the divergence-free field ${\bf B}$ and the function
$\varphi$ satisfy
\begin{equation}
    {\partial {\bf B} \over \partial t} - 
    \nabla \times ( {\bf u} \times {\bf B} ) =0 \;,\;\;\;
    {\partial \varphi \over \partial t} +
     {\bf u} \cdot \nabla \varphi =0     \label{beq}
\end{equation}
which are the frozen field equations. Then

(1) $\partial_{t}+u$ is a Hamiltonian vector field with the symplectic
two-form
\begin{equation}
  \Omega = - (\nabla \varphi + {\bf u} \times {\bf B}) 
          \cdot d{\bf x} \wedge dt + {\bf B} \cdot 
             (d{\bf x} \wedge d{\bf x})    \label{symp2}
\end{equation}
and the Hamiltonian function $\varphi$.

(2) $\rho_{\varphi} \equiv - {\bf B} \cdot \nabla \varphi$ is the invariant
Liouville volume density. 

(3) If moreover ${\bf B} = \nabla \times {\bf A}$ for some vector 
potential ${\bf A}$, then $\Omega$ is exact
\begin{equation}
 \Omega = -d \theta \;,\;\;\; 
        - \theta =  \psi \, dt + {\bf A} \cdot d{\bf x}     \label{cone2}
\end{equation}
where $\psi$ is determined by the equation 
\begin{equation}
  {\partial {\bf A} \over \partial t} -
  {\bf u} \times (\nabla \times {\bf A}) = 
  \nabla (\varphi + \psi)     \;.            \label{aeq}
\end{equation}

(4) The Darboux coordinates for $\Omega$ defines the Clebsch
representation of the vector potential ${\bf A}$.
\end{theorem}

We shall now show with an example from dynamo theory that 
the Hamiltonian character of Lagrangian description is determined by 
dynamics of the ${\bf B}$ field rather than the velocity itself.
The dynamics of the velocity field 
is determined by the so-called magneto-convection equation (\ref{geuler})
where the complicated velocity dependent force field per unit mass
\begin{equation}
   {\bf F}= - {1 \over \rho} {\bf B} \times (\nabla \times {\bf B}) 
            + \nu \nabla^{2} {\bf u} - 2 {\bf \Omega} \times {\bf u}
            + C {\bf g} - \pi {\bf u}
\end{equation}
includes Lorentz, viscous, coriolis, gravity forces and the reduced 
pressure. We refer to \cite{robert} for a description of 
the physical framework and terminology involved.
The divergence free magnetic field satisfies
\begin{equation}
    {\partial {\bf B} \over \partial t} - 
    \nabla \times ( {\bf u} \times {\bf B} ) = \eta \nabla^{2} {\bf B}
\label{dbeq}      \end{equation}
where $\eta$ is the magnetic diffusivity. With this equation the two-form
\begin{equation}
  \Omega_{d} = - (\nabla \varphi + {\bf u} \times {\bf B} 
        - \eta \nabla \times {\bf B} ) \cdot d{\bf x} \wedge dt 
        + {\bf B} \cdot (d{\bf x} \wedge d{\bf x})    \label{sympd}
\end{equation}
is symplectic. It is exact with the one-form (\ref{cone2}) where
the scalar potential $\psi$ is determined from eq(\ref{aeq}) whose
right hand side is replaced by $\nabla (\psi + \varphi 
- \eta \nabla \cdot {\bf A}) + \eta \nabla^{2} {\bf A}$.
However, the existence of a Hamiltonian function is prevented
by the diffusion term in eq(\ref{dbeq}). For $\eta =0$, or equivalently
for infinite conductivity, the suspension of velocity field is
Hamiltonian regardless of how complicated the force field may be.

\section{Integral Invariants and Eulerian Conservation Laws}

Conservation laws of dynamical equations are
divergence expressions of the form
\begin{equation}
     {\partial T  \over \partial t} + \nabla \cdot {\bf P} = 0  \label{cola} 
\end{equation}     
where the conserved density $T$ and the flux ${\bf P}$ are functions
of $t,{\bf x}$, the Eulerian fields and their derivatives.
Some of these Eulerian conservation laws are known to be related 
to the Lagrangian conserved quantities. 
In this section we shall use the symplectic structure to construct
invariant differential forms of the Lagrangian picture and show with
examples their connection with conservation laws of the Eulerian description.

We start with the observation that eq(\ref{cola})
can be expressed as the closure of a three-form in  
four-dimensional space-time. If $\Theta$ is the closed three-form
corresponding to eq(\ref{cola}), then we have for its Lie derivative
\begin{equation}
      \pounds_{\partial_{t}+u} ( \Theta )= d \Sigma  \label{inv}
\end{equation}
where $\Sigma = i(\partial_{t}+u)(\Theta)$. This
means that $\Theta$ is a relative integral invariant of $\partial_{t}+u$.
Conversely, it turns out that some conserved densities of the Eulerian
dynamics can be constructed out of invariant forms of symplectic
structure of Lagrangian kinematics. 
The formulation of the continuity equation (\ref{divo}) as the closure 
of invariant three-form $\Theta_{M}= \rho \prod_{i=1}^{3}(dx^{i}-u^{i}dt)$
can be viewed as a primitive example of the sort of relations we wish to
consider.

\subsection{Helicity Conservation}

The canonical one-form $\theta$ is 
a relative integral invariant for its Lie derivative
$\pounds_{\partial_{t}+u} ( \theta)=
      d(i(\partial_{t}+u) (\theta) - \varphi )$
is only closed. Its exterior derivative which is the symplectic
two-form is then an absolute invariant.
Using them we construct the (relative) invariant three-form 
\begin{eqnarray}
 \Theta_{\cal H} &=& \theta \wedge d \theta \, = \,
 {\bf A} \cdot \nabla \times {\bf A} \, dx \wedge dy \wedge dz + \nonumber \\
       & & \;\;\;\; ( (\psi + {\bf u} \cdot  {\bf A}) {\bf B}
        - ( {\bf A} \cdot  {\bf B}) {\bf u}
        -  {\bf A} \times  \nabla \varphi  )  \cdot
       (d{\bf x} \wedge d{\bf x}) \wedge dt 
\end{eqnarray}
which consists of the well-known (magnetic) helicity and corresponding
flux densities. The identity $d (\theta \wedge d \theta ) 
- \Omega \wedge \Omega =0$
expressing the fact that its closure is an absolute invariant,
gives the conservation law
\begin{equation}
 { \partial \over \partial t} ( {\bf A} \cdot {\bf B} )
       + \nabla \cdot ( ( {\bf A} \cdot  {\bf B}) {\bf u} 
  - (\psi + \varphi + {\bf u} \cdot  {\bf A}) {\bf B} ) = 0    \label{hel}
\end{equation}
for the dynamics governed by general force fields. In the
case of Euler flow eq(\ref{hel}) is the conservation law 
for the helicity ${\bf u} \cdot {\bf w}$ which is the Casimir
of the Lie-Poisson bracket for the Euler equations in the vorticity
variable$^{\cite{mr}}$.

\subsection{Energy Conservation}

The energy conservation arises from a somewhat different consideration.
If we regard the closure conditions of symplectic two-form
$\Omega$ as the homogeneous part of Maxwell equations, the
inhomogeneous part will define the exact three-form
\begin{equation}
  \Theta_{E} =  q \, dx \wedge dy \wedge dz 
       - {\bf J} \cdot (d{\bf x} \wedge d{\bf x}) \wedge dt 
\end{equation}
where the expressions for charge and current densities
\begin{equation}
     \nabla \cdot {\bf E} = q  \;,\;\;\;
  - {\partial {\bf E} \over \partial t} 
  + \nabla \times {\bf B} = {\bf J}      \label{nonmax}
\end{equation}
are the usual form of inhomogeneous Maxwell equations.
From $\Omega$ we read off the electric field
${\bf E} = - ( \nabla \varphi + {\bf u} \times {\bf B})$.
The closure of $\Theta_{E}$ is equivalent to the well-known 
Eulerian conservation law
\begin{equation}
  d \Theta_{E} =({\partial q \over \partial t} + \nabla \cdot {\bf J})
                  \;dt \wedge dx \wedge dy \wedge dz = 0   \label{chcon}
\end{equation}
which easily follows from definitions (\ref{nonmax}).
For the symplectic structure (\ref{symp2}) the conserved density $q$ 
is given by 
\begin{equation}
  q = - \nabla^{2} \varphi - {\bf \omega} \cdot {\bf B} + 
             {\bf u} \cdot ( \nabla \times {\bf B})  
\end{equation}             
which reduces to $\nabla^{2} \psi$ in the Coulomb 
(or transverse) gauge $\nabla \cdot {\bf A}=0$. Note that this
gauge condition is satisfied automatically for $\theta_{e}$.
In the case of Euler equation we have
\begin{equation}
  q_{e} = - \nabla^{2} h -  \omega^{2} + 
             {\bf u} \cdot ( \nabla \times {\bf \omega})    
       = - \nabla^{2} ( h+{1 \over \rho}P+{1 \over 2}u^{2}) 
       = \nabla^{2} \psi_{e}
\end{equation}             
which is the form of energy density analogous to the one
considered in \cite{hegna} for the case of steady flows.

\section{An Infinite Family of Invariants}

The introduction of $\Theta_{E}$ is more crucial for the present 
section than for the relatively trivial conservation law (\ref{chcon}). 
It naturally exists via definition (\ref{nonmax}) for any
hydrodynamical system having one vectorial frozen-in field.
One of the obvious exceptions is the viscous flow of
the Navier-Stokes equation, and another is the fluid with finite
conductivity which we have discussed before.

\subsection{A New Symmetry and Associated Invariants}

We shall show that the requirement on $\Theta_{E}$ to be an 
absolute invariant results in a new symmetry of the velocity field,
gives an Eulerian conservation law and turns $q$ into a conserved
density of the Lagrangian description as well.

\begin{theorem}
$\Theta_{E}$ is an absolute invariant of $\partial_{t}+u$ if and
only if the divergence-free vector field $q{\bf u}-{\bf J}$ is
a time-dependent symmetry of ${\bf u}$. In this case, 

(1) $q$ is also a Lagrangian invariant and

(2) ${\bf A} \cdot (q{\bf u}- {\bf J})$ is an Eulerian conserved density 
depending on dynamics of velocity field through the definition of ${\bf J}$.
\end{theorem}
{\bf Proof:} Since the Lie derivative of $\Theta_{E}$ 
\begin{equation}
      \pounds_{\partial_{t}+u} ( \Theta_{E} ) =
       d ((q {\bf u}-{\bf J}) \cdot  (d{\bf x} \wedge d{\bf x})
      - ({\bf J} \times {\bf u}) \cdot d{\bf x} \wedge dt)   \label{dmx3}
\end{equation}
is exact, it is in general a relative invariant.
The condition that $\Theta_{E} $ is an absolute invariant is the
vanishing of the right hand side of (\ref{dmx3}). This implies that
$q{\bf u}-{\bf J}$ must be divergence-free and satisfy
\begin{equation}
  {\partial (qu-J) \over \partial t} + [ u,qu-J ]
   = - (\nabla \cdot {\bf u}) (qu-J)                 \label{symuj}
\end{equation}
which is the equation defining a symmetry of $u$.
In particular, given $u$ and $q$ for an incompressible flow, 
the solutions $J$ of linear inhomogeneous first order equations
\begin{equation}
  {\partial J \over \partial t} + [ u,J ]
   = q {\partial u \over \partial t} \label{syuj}
\end{equation}
makes $qu-J$ into a symmetry of $u$.

The invariance of $q$ under the unsteady flow of $u$ 
\begin{equation}
   { {\partial q} \over \partial t} + {\bf u} \cdot \nabla q =0   
\end{equation}
follows from the divergence-free condition and 
the conservation law (\ref{chcon}).

To prove (2), we first observe that
if $\Theta_{E}$ is an absolute invariant, so is the closed two-form 
$\Sigma_{E} \equiv i(\partial_{t}+u) ( \Theta_{E} )$.
Then the three-form $\theta \wedge \Sigma_{E}$ is a relative invariant and 
the conservation law
\begin{equation}
  {\partial \over \partial t}
      ({\bf A} \cdot (q{\bf u}- {\bf J}))
   + \nabla \cdot ({\bf A} \times ({\bf J} \times {\bf u})   
   - (\varphi + \psi ) (q{\bf u}- {\bf J})) =0            \label{cons}
\end{equation}
follows from the identity
$d(\theta \wedge \Sigma_{E})+ \Omega \wedge \Sigma_{E}=0$. $\bullet$

\subsection{New Symmetries and Underlying Geometry}

It can easily be shown that the Lie bracket of symmetries
${\bf B}$ (or ${\bf w}$) and $q{\bf u}-{\bf J}$ of ${\bf u}$
is a new time-dependent symmetry. In fact, this is true for any two 
symmetries of $u$.
Thus, repeated applications of the Lie bracket generates
an infinite dimensional algebra of time-dependent symmetries
of the velocity field. 

New integral invariants can then be constructed from Lie derivatives
and inner products of known differential invariants with these 
symmetries (see for example, exercise 6.18 of ref \cite{olver}). 
Relations with Eulerian conservation laws are established by taking
appropriate products and combinations of invariant forms. The derivations
of conservation laws (\ref{hel}) and (\ref{cons}) are examples
of this construction.

To this end, we wish to obtain a geometric characterization of
the origin of conservation laws generated by the above procedure. 
In order to do this, we shall first restrict
them to a particular gauge in which the densities are also
conserved along the trajectories of velocity field. The resulting 
Lagrangian conservation laws will be familiar from geometry.
The appropriate gauge
can be characterized by demanding the canonical one-form 
$\theta$ to be an absolute invariant of $\partial_{t}+u$,
that is, by the condition $ \varphi + \psi + {\bf u} \cdot {\bf A} 
= constant $.
Then for an incompressible fluid, the conserved densities in 
eqs(\ref{hel}) and (\ref{cons}), namely the magnetic helicity 
${\bf A} \cdot {\bf B}$, and ${\bf A} \cdot (q{\bf u}- {\bf J})$ 
are also Lagrangian invariants. 

We shall now show by further restricting to Euler equation that  
the infinite family of invariants have
a geometric (rather than dynamic) origin in the description of
Euler flow as a geodesic motion on an infinite dimensional group.
For the Euler equation (\ref{euler}) the gauge condition implies 
that $\theta_{e}$ is absolutely invariant on surfaces defined by 
$u^{2}/2-P(\rho)/ \rho = constant$. The corresponding Lagrangian
invariants are the helicity ${\bf u} \cdot {\bf w}$, and 
${\bf u} \cdot (q{\bf u}- {\bf J})$. 
The restriction on dynamics for which the velocity field admits
this sort of invariants is contained in the following
straightforward result.
\begin{theorem}
Let ${\bf v}$ be a time-dependent symmetry of velocity field
${\bf u}$ whose dynamics is governed by eq(\ref{geuler}). If
\begin{equation}
      {\bf v} \cdot ( {\bf F} + {1 \over 2} \nabla u^{2})=0 \;
\end{equation}
then ${\bf u} \cdot {\bf v}$ is a Lagrangian invariant.
\end{theorem}
This explains the relevance of the chosen gauge for the Euler equation.
As an immediate consequence, there follows the conservation of 
${\bf u} \cdot {\bf B}$ in a magnetic relaxation problem which was 
also indicated in \cite{kuzmin}.
Note that there is no dynamical restriction for invariants of 
the form ${\bf A} \cdot {\bf v}$.

The Lagrangian conservation laws which are in the form of a product
of velocity with a symmetry are familiar from
geometry where one proves, as for example in \cite{mtw},
that if $u$ is tangent to a 
geodesic curve on a Riemannian manifold and if $v$ is 
a Killing vector (an infinitesimal isometry), then 
${\bf u} \cdot {\bf v}$ is an invariant
of any motion with velocity $u$. 
The analogy with the above invariants of Lagrangian
description will become transparent if we recall that
in Lagrangian coordinates the Euler flow is a geodesic motion
on the infinite dimensional group of volume preserving 
diffeomorphisms$^{\cite{via},\cite{mr}}$.

Since $u_{t}$ is right invariant (c.f. eq(\ref{vel})), the motion
is generated by left translation.
The symmetry criterion for $v_{t}(x)=v(t,x)$ is the infinitesimal 
expression for the invariance of $v_{t}$ under the pull-back 
$g_{s}^{*}(v_{t}) \equiv Tg_{s}^{-1} \circ v_{t} \circ g_{s}$ by geodesic
motion. So, any symmetry generator is a left invariant vector field.
Moreover, since the starting symmetries $w$ and $qu- J$ are 
divergence-free, it follows from the identity
\begin{equation}
   div( [ v,w ] ) = v \,div(w) - w \, div(v)  
\end{equation}
that they all are divergence-free and hence are elements of the algebra
of volume preserving diffeomorhisms. The left invariance of symmetries
implies that they are the generators of the right action or, in fluid
mechanical context, the particle relabelling symmetry.
Thus, the conserved densities we have been considered so far are all
associated with this general symmetry property of the Euler flow.
It has indeed been shown explicitly
(see \cite{salmon} and the references therein)
that many of the well-known conserved densities of fluid motion are
derivable from a so-called general vorticity conservation law 
connected with the particle relabelling symmetry.
We may therefore claim that
the infinite dimensional symmetry algebra constructed above is 
the left Lie algebra of group of volume preserving diffeomorphisms
and it consists of infinitesimal generators of the particle 
relabelling symmetry.

\section{Discussion and Conclusions}

We have shown that the symplectic structure of kinematic description  
presents an interplay with dynamical description.
The connection between invariants of velocity field and
conservation laws of dynamical equations is another
manifestation of this interplay. 

We obtained the Eulerian dynamical equations (\ref{weq}), (\ref{beq}) 
and (\ref{aeq}) from a purely geometric 
treatment inherent in the Lagrangian
description. In an earlier work, described in \cite{kozlov},
Kozlov obtained eqs(\ref{reuler}) and (\ref{weq}) in constructing integral
manifolds of finite dimensional non-autonomous dynamical systems
thereby first recognizing the connection with hydrodynamics of an
ideal fluid. Motivated by physical arguments Kuz'min introduced in
\cite{kuzmin} (see also \cite{sagdeev}) fields
satisfying (in a particular gauge) eq(\ref{aeq}) and used them
to construct Lagrangian invariants.

In this work we presented a much more suitable construction
relying on symplectic geometry for a systematic and exhaustive 
investigation of both Lagrangian and Eulerian invariants. 
Moreover, the construction inherits geometric settings not only
for the invariants themselves but also for their defining equations.
We therefore conclude that the Lagrangian description, 
which is seldom used in fluid dynamics, provided with a 
geometrical framework, may become a powerful tool in analysis 
and the construction of invariants of hydrodynamical models.

\section*{Acknowledgement}
I am indepted to Professors Ay\c{s}e Erzan and Alp Eden 
for insisting on me to participate to a workshop
on vortex dynamics in \.Istanbul where I took the initiative 
for this work. I acknowledge the illuminating discussions and
encouragements of Prof. Erzan,  Prof. Yavuz Nutku and
Prof. Cihan Sa\c{c}l{\i}o\u{g}lu.

\end{document}